\documentclass[english]{article}
\usepackage[T1]{fontenc}
\usepackage[latin1]{inputenc}

\makeatletter

\newcommand{\lyxaddress}[1]{
\par {\raggedright #1
\vspace{1.4em}
\noindent\par}
}

\usepackage{babel}
\makeatother
\begin{document}

\title{Revisiting Boole Equation in the Quantum Context}

\author{Rudolf Muradian and Diego Frias}

\date{}

\maketitle

\lyxaddress{\begin{center}
\emph{CEPEDI, Centro de Pesquisa em Informática de Ilhéus, Depatamento
de Computacão Quântica, Ilhéus BA, Brazil }
\par\end{center}}

\lyxaddress{\begin{center}
\emph{UESC, Universidade Estadual de Santa Cruz, Departamento de
Ciências Exatas e Technologicas, Ilhéus BA, Brazil }
\par\end{center}}

\begin{abstract}
In this work we try to clarify the fundamental relationship between
\emph{bits} and \emph{qubit}s, starting from very simple George Boole
equation. We derive a generic and compact expression for basis vectors
of \emph{qubit} which can be useful in a further application. We also
derive a generic form for the projection operator in the quantum information
space. The results are also extended to higher dimensional \emph{d}-level
cases of \emph{qutrits} and \emph{qudits}.
\end{abstract}

\section{Introduction\label{sec:Introduction}}

Quantum information theory is related to the most fundamental aspects
of the computer science. In this work we investigate the transition
from classical to quantum information. We propose a framework for
understanding the relationship between \emph{bit} and \emph{qubit}
based on Boole equation $x^{2}=x$ . The same procedure is then applied
for devising a general expression for the basis vectors of the \emph{d-}level
quantum information unit \emph{qudit}, as in (\ref{eq:qudit}). In
fact, we demonstrate that the elements of orthonormal basis in \emph{d-}dimensional
Hilbert space $C^{d}$ can be represented in a very simple generic
form.

\section{Bit versus Qubit\label{sec:Bit-versus-Qubit}}

The elementary unit of information in classical computation is the
\emph{Shannon bit} or simply \emph{bit}, which can take only two values
$x\in\{0,1\}$ . Shannon bit was introduced under straightforward
influence of the ideas of the great mathematician and thinker George
Boole exposed in his celebrated book \cite{boole}. Here, on the page
22 Boole introduced his famous equation 

\begin{equation}
x^{2}=x\label{eq:boole}\end{equation}
and continued on the page 26: 

\begin{quote}
{}``We have seen ... that the symbols of Logic are subjects to the
special law $x^{2}=x$. Now of the symbols of Number there are but
two, viz. 0 and 1, which are subject to the same formal law . We know
that $0^{2}=0$ and that $1^{2}=1$; and the equation, considered
as algebraic, has no other roots than 0 and 1.''
\end{quote}
We are going to demonstrate that qubit, a quantum generalization of
bit, could be deduced from the particular matrix generalization of
the same equation, namely\begin{equation}
P(x)^{2}=P(x),\,\quad x\in\{0,1\}\label{eq:projeq}\end{equation}
where $x$ is the solution of the usual Boole equation (1). Equation
(\ref{eq:projeq})with normalization condition $TrP(x)=1$ can be
solved analytically to give \begin{equation}
P(x)=\left(\begin{array}{cc}
1-x & 0\\
0 & x\end{array}\right)\label{eq:proj}\end{equation}

It is straightforward to see this using identities $x^{2}=x$ and
$(1-x)^{2}=1-x$, which holds for any $x\in\{0,1\}$ . The solution
(\ref{eq:projeq}) corresponds to the \emph{projection operator} (\emph{state
operator or filter operator}) in Quantum Mechanics \cite{dirac},
\cite{peres} and in general any projection operator \emph{P} has
the property $P^{2}=P$. Projection operator \emph{P(x)} from (\ref{eq:proj})
can be represented in terms of Dirac's \emph{kets \begin{equation}
\mid x>=\left(\begin{array}{c}
1-x\\
x\end{array}\right)\label{eq:ket}\end{equation}
}and \emph{bras}

\begin{equation}
<x\mid=\left(\begin{array}{cc}
1-x & x\end{array}\right)\label{eq:bra}\end{equation}
as outer $ket\otimes bra$ product\begin{equation}
P(x)=\mid x><x\mid\label{eq:projketbra}\end{equation}

From the definition (\ref{eq:ket}) it follows the familiar form of
two basis vectors $\mid0>=\left(\begin{array}{c}
1\\
0\end{array}\right)$and $\mid1>=\left(\begin{array}{c}
0\\
1\end{array}\right)$, which confirms that Dirac's \emph{kets} $\mid0>$and $\mid1>$ are
quantum generalizations of Boolean 0 and 1.

\section{Extension to Qutrit and Qudit\label{sec:Extension-to-Qutrit}}

\subsection{Qutrit\label{sub:Qutrit}}

The amount of information in 3-level (ternary) classical system is
named \emph{trit} and can assume three values, such as $x\in\{ yes,\, no,\, unknown\}$
or $x\in\{0,\,1,\,2\}$ . Similarly the unit of quantum information
in 3-level quantum system (e. g. spin 1 particle in magnetic field)
is called \emph{quantum trit} or \emph{qutrit}. We shall show that
the appropriate $3\times3$ solution of matrix generalization of the
Boole equation can be used to introduce 3-dimensional normalized \emph{qutrit}
ket vector in compact form\begin{equation}
\mid x>=\frac{1}{2}\left(\begin{array}{c}
(1-x)(2-x)\\
2x(2-x)\\
x(x-1)\end{array}\right),\,\,\,\,\,\, x\in\{0,1,2\}\label{eq:qutrit}\end{equation}

The classical Boole equation for a \emph{trit} is a cubic equation\begin{equation}
x(x-1)(x-2)=0\label{eq:booletrit}\end{equation}

The corresponding quantum matrix equation \begin{equation}
P(x)^{2}=P(x),\,\quad x\in\{0,1,2\}\label{eq:proj3}\end{equation}
has $3\times3$ matrix solution in the form\begin{equation}
P(x)=\frac{1}{2}\left(\begin{array}{ccc}
(1-x)(2-x) & 0 & 0\\
0 & 2x(2-x) & 0\\
0 & 0 & x(x-1)\end{array}\right)\label{eq:projQutrit}\end{equation}
which can be also expressed as outer product $P(x)=\mid x><x\mid$
with $\mid x>$ given as in (\ref{eq:qutrit}). From (\ref{eq:qutrit})and
(\ref{eq:projQutrit}) we can see that qutrit is described by the
following three basis vectors\begin{equation}
\mid0>=\left(\begin{array}{c}
1\\
0\\
0\end{array}\right),\,\,\mid1>=\left(\begin{array}{c}
0\\
1\\
0\end{array}\right),\,\,\mid2>=\left(\begin{array}{c}
0\\
0\\
1\end{array}\right),\,\,\label{eq:qitrit}\end{equation}
with corresponding projection operators\begin{equation}
P(0)=\left(\begin{array}{ccc}
1 & 0 & 0\\
0 & 0 & 0\\
0 & 0 & 0\end{array}\right),\,\, P(1)=\left(\begin{array}{ccc}
0 & 0 & 0\\
0 & 1 & 0\\
0 & 0 & 0\end{array}\right),\,\, P(2)=\left(\begin{array}{ccc}
0 & 0 & 0\\
0 & 0 & 0\\
0 & 0 & 1\end{array}\right),\,\,\label{eq:qutritProj}\end{equation}

\subsection{Qudit\label{sub:Qudit}}

A unit of quantum information in \emph{d}-level quantum system, \emph{qudit,}
can be introduced in the same manner. For simplicity, let us consider
first the particular case \emph{}for \emph{d}=4. The same considerations
allow to obtain the basis \emph{ket} vectors of 4-level system as\begin{equation}
\mid x>=\frac{1}{6}\left(\begin{array}{c}
(1-x)(2-x)(3-x)\\
3x(2-x)(3-x)\\
3x(x-1)(3-x)\\
x(x-1)(2-x)\end{array}\right)\,\,\,\, x\in\{0,1,2,3\}\label{eq:ket4}\end{equation}
and the corresponding $4\times4$ projection operator \emph{P(x)}
which has only non-null diagonal terms defined by entries of the vector
above, that is \begin{equation}
P(x)=\frac{1}{6}\, diag((1-x)(2-x)(3-x),\,3x(2-x)(3-x),\,3x(x-1)(3-x),\, x(x-1)(2-x))\label{eq:proj4}\end{equation}

This projection operator is a solution of matrix Boole equation for
$x\in\{0,1,2,3\}$ with properties\begin{equation}
P(x)^{2}=P(x),\,\,\,\,\, TrP(x)=1\label{eq:prTr}\end{equation}

The completeness relation is fulfilled:\begin{equation}
{\displaystyle {\displaystyle {\textstyle {\displaystyle \sum_{x=0}^{3}}}}}P(x)=\left(\begin{array}{cccc}
1 & 0 & 0 & 0\\
0 & 1 & 0 & 0\\
0 & 0 & 1 & 0\\
0 & 0 & 0 & 1\end{array}\right)\label{eq:complet}\end{equation}

It is easy to generalize this results to the case of general \emph{qudit}.
For \emph{d}-level system $x\in\{0,1,2,...,d-1\}$ the basis ket vector
takes form\begin{equation}
\mid x>=\left(\begin{array}{c}
\frac{(1-x)(2-x)\cdots\cdots\cdots(d-1-x)}{(d-1)!}\\
\frac{(0-x)(2-x)\cdots\cdots\cdots(d-1-x)}{(d-2)!}\\
\vdots\\
\frac{1}{(-1)^{k}k!((d-1-k)!)}\prod_{k^{'}=0,\, k^{'}\neq k}^{d-1}(k^{'}-x)\\
\vdots\\
\frac{(0-x)(1-x)\cdots\cdots\cdots(d-2-x)}{(-1)^{d-1}(d-1)!}\end{array}\right)\,\,\,\, x\in\{0,1,2,...,d-1\}\label{eq:qudit}\end{equation}

A general normalized d-dimensional vector can be expanded in this
basis as\begin{equation}
\sum_{x=0}^{d-1}a_{x}\mid x>\label{eq:norm}\end{equation}
where $a_{x}$ are complex numbers satisfying $\sum_{x}|a_{x}|^{2}=1$.

\section*{4 Example\label{sec:4-Example}}

The representation of qubit basis vectors in the form (\ref{eq:ket})
allows to represent the entangled Bell basis for two qubits in a compact
form. Using well known circuit constructed from \emph{Hadamard} and
\emph{cnot} gates, we obtain from (\ref{eq:ket}) the following explicit
compact form for Bell states\begin{equation}
|B_{xy}>=\frac{1}{\sqrt{2}}\left(\begin{array}{cccc}
1 & 0 & 1 & 0\\
0 & 1 & 0 & 1\\
0 & 1 & 0 & -1\\
1 & 0 & -1 & 0\end{array}\right)\left(\begin{array}{c}
(1-x)(1-y)\\
y-xy\\
x-xy\\
xy\end{array}\right)=\frac{1}{\sqrt{2}}\left(\begin{array}{c}
1-y\\
y\\
y-2xy\\
(1-2x)(1-y)\end{array}\right)\label{eq:bell}\end{equation}
which can be compared with other well known expressions, e. g. in
\cite{nielsen}.

\section*{5 Conclusions\label{sec:5-Conclusions}}

Starting from a rather simple Boole equation and its matrix generalization
we have devised a generic and compact representation of basis vectors
for \emph{qubit,} \emph{qutrit} and general \emph{qudit} case. We
hope that our results could help in formalizing quantum algorithms
and utilized by the researcher in the field of quantum information
and computing.

\paragraph{Acknowledgment}

This work was supported by FAPESB (Research Foundation of the Bahia
State, Brazil).

\end{document}